\documentclass[prb,aps,reprint,twocolumn,showpacs, amsmath,amssymb,superscriptaddress]{revtex4-1}
\usepackage{graphicx,color}% Include figure files
\usepackage{dcolumn}% align table columns on decimal point
\usepackage{float}
\usepackage{placeins}
\usepackage{wrapfig, blindtext}
\usepackage{booktabs,siunitx}
\usepackage{multirow,tabularx,booktabs} % JEB : for nicer table
\usepackage[titletoc,toc,title]{appendix}
\usepackage{pst-node,lipsum}
\usepackage[caption=false]{subfig}
\usepackage[font=small,labelfont=bf,tableposition=top]{caption}
\usepackage[compact]{titlesec}         % you need this package
\titlespacing{\section}{0pt}{0pt}{0pt} % this reduces space between (sub)sections to 0pt, for example
\AtBeginDocument{%                     % this will reduce spaces between parts (above and below) of texts within a (sub)section to 0pt, for example - like between an 'eqnarray' and text
	\setlength\abovedisplayskip{0pt}
	\setlength\belowdisplayskip{0pt}}

\begin{document}
	
	%\preprint{APS/123-QED}
	%\lipsum[1-2]
	\title{Re-thinking CO adsorption on transition-metal surfaces: Density-driven error ?}
	
	\author{Abhirup Patra}
	\affiliation{Department of Physics, Temple University, Philadelphia, PA 19122}
	\affiliation{School of Materials Science and Engineering, Georgia Institute of Technology, Atlanta, GA 30308}
	\author{Haowei Peng}
	\affiliation{Department of Physics, Temple University, Philadelphia, PA 19122}
	\author{Jianwei Sun}
	\affiliation{Department of Physics and Engineering Physics, Tulane University, New Orleans, LA 70118}
	\author{John P. Perdew}
	\affiliation{Departments of Physics and Chemistry, Temple University, Philadelphia, PA 19122}
	%    \affiliation{ Department of Chemistry,Temple University, Philadelphia,PA-19122}
	\date{\today}% It is always \today, today,
	
	\begin{abstract}
	Adsorption of the molecule CO on metallic surfaces is an important unsolved problem in Kohn-Sham density functional theory (KS-DFT). We present a detailed study of carbon monoxide adsorption on fcc (111) surfaces of 3d, 4d and 5d metals using nonempirical semilocal density functionals for the exchange-correlation energy: the local-density approximation (LDA), two generalized gradient approximations or GGAs (PBE and PBEsol), and a meta-GGA (SCAN). The typical error pattern (as found earlier for free molecules and for free transition metal surfaces), in which results improve from LDA to PBE or PBEsol to SCAN, due to the satisfaction of more exact constraints, is not found here. Instead, for CO adsorption on transition metal surfaces, we find that, while SCAN overbinds much less than LDA, it overbinds slightly more than PBE. Moreover, the tested functionals often predict the wrong adsorption site, as first pointed out for LDA and GGA in the “CO/Pt (111) puzzle”. This abnormal pattern leads us to suspect that the errors of PBE and SCAN for this problem are density-driven self-interaction errors associated with incorrect charge transfer between molecule and metal surface. We point out that, by the variational principle, overbinding by an approximate functional would be reduced if that functional were applied not to its selfconsistent density for the adsorbed system but to an exact or more correct density for that system. Finally, we show for CO on Pt(111) that the site preference is corrected and the adsorption energy is improved for the PBE functional by using not the selfconsistent PBE density but a PBE+U density. The resulting correction to the PBE total energy is much larger for the adsorbed system than for its desorbed components, showing that the error is in the density of the adsorbed system. This seems to solve the Feibelman 2001 CO/Pt(111) puzzle, in principle if not fully in practice. 
	\end{abstract}
	
	\maketitle
	\vspace*{0.5cm}
	\section{\label{sec:intro}INTRODUCTION}
	\vspace*{0.5cm}
	Adsorption of inorganic and organic molecules on different surfaces is an important problem in surface science \citep{norskov2011density,catalysis-1,catalysis-4}, both in theory and in experiment. The adsorption of CO on metallic surfaces is a well-known example\citep{HammerPRL96,DOYENSURFSC74}. For the past few decades, much effort has been devoted to studying the adsorption of CO on transition metals, which is relevant to the catalytic oxidation of CO in industry. This adsorption is considered as a “prototype” that mimics many other interesting and practically important adsorption processes. For the cases considered here, the CO molecule stands straight up on the metal surface, with its carbon atom closer to the metal atoms.

	Modern electronic structure theory, especially density functional theory (DFT) \citep{KOHN-DFT1,KOHN-DFT2}, is widely used to describe many surface-related problems including molecular adsorption. However, in 2001 Feibelman et al.\cite{CO-Pt-Puzzle-1} challenged the accuracy of GGA and LDA functionals for CO adsorption on the Pt (111) surface. Their work showed that GGA and LDA are both qualitatively and quantitatively wrong in their predictions of the adsorption site of CO on the Pt (111) surface, independent of the technical details of the calculation. This study presented a ``CO/Pt (111) puzzle", which has been investigated\citep{CO-Pt-Puzzle-2,CO-Pt-Puzzle-3} further by many electronic structure theory methods including different levels of approximations within Kohn-Sham-DFT (KS-DFT).  By now, all five rungs of Jacob's ladder\cite{MRS-JOHN,jacobsladder} of density functional approximations have been used to study this particular problem. The non-empirical 
	functionals LDA\cite{KOHN-DFT2}, GGAs (PBE, PBEsol)\citep{PBE,perdew2008restoring} and meta-GGAs (TPSS, revTPPS)\citep{TPSS,revTPSS} fail to agree with the picture of CO adsorption from low-energy electron diffraction (LEED) and electron energy loss spectroscopy (EELS). Many previous studies showed that the hollow adsorption site is preferred by these semilocal approximations, while the low-coordination top site is preferred in the experiments. A recent study by Janthon et al.\cite{CO-Pt-Puzzle-vdW} reported that the semi-empirical M06-L meta-GGA predicts both correct adsorption site and adsorption energy. Sun et al.\cite{sun2011improved}showed that a revised version (revTPSS) of the TPSS meta-GGA significantly improves the surface energies and adsorption energies for transition metals. These two studies suggest that inclusion of the kinetic energy density ($\tau$), the added ingredient that defines a meta-GGA, can be important for the surface properties of metals. A recent study\cite{PatraPNAS2017} shows that inclusion of a vdW correction\cite{rVV10} to the nonempirical SCAN meta-GGA\cite{SCAN} yields accurate surface energies and work functions of free transition-metal surfaces. SCAN+rVV10 also correctly predicts\cite{rVV10} chemisorption and physisorption minima in the binding of graphene to Ni(111), with the physisorption minimum in good agreement with RPA calculations\cite{OT13}. Here we will investigate whether constraint-satisfying nonempirical meta-GGAs such as SCAN can correctly describe CO adsorption on transition-metal surfaces, and, if not, why not. 
	
	Semilocal functionals that satisfy more physical constraints are sometimes considered to be closer than the exact functional\cite{JohnScience}, and SCAN\cite{SCAN} was constructed to satisfy all 17 exact constraints that a meta-GGA can. SCAN is considerably more accurate than PBE for molecules\cite{GoerigkPCCP2017,HH} and for condensed matter (defects in semiconductors \cite{SR16}, structural phase transitions of solids under pressure\cite{CHANDRAPRB}, ferroelectrics\cite{YUBO2017}, ice\cite{SR16}, liquid water\cite{SCANWATERPNAS}, formation energies and ground-state crystal structures of solids\cite{JohnNPJComp}, cuprates\cite{JianweiCommunication18}, etc.).  But the exact constraints embedded in SCAN can have a more directly beneficial effect upon the energy for a given density than upon the selfconsistent density, especially where there are charge transfer errors\cite{PPLB,RPC} due to self-interaction\cite{PZSIC}.  
	
	The energy error of an approximate functional is the sum of a functional error (the error that it makes when applied to the exact density) and a density-driven error (the difference between the energies of the approximate functional applied to its selfconsistent density and to the exact density)\cite{KeironPRL2013}. Although in most cases the functional error is larger than the density-driven error, the latter by itself can produce a non-physical result\cite{KeironPRL2013,KeironJCP2014,SKSBHB}. For realistic bonding situations, a good semilocal functional like SCAN could have a small functional error but still have noticeable density-driven error. In other words, even the best semilocal functional can have a qualitatively-wrong functional derivative. We suspect that this is the case for PBE and SCAN applied to molecules chemisorbed on metal surfaces.
	
	The most serious density-driven errors are charge-transfer errors, which are directly relevant to the strong binding of a closed-shell molecule like CO to a metal surface. The Kohn-Sham molecular orbitals of the free molecule can evolve in the adsorbed molecule into energy-broadened resonances of the bulk metallic orbitals. In Blyholder's model\cite{Blyholder-model}, bonding of CO on metal surfaces can be described by $\sigma$  bonding through electron transfer from the filled $5\sigma$ (HOMO) orbital of CO to an unfilled \textit{d}-orbital of the metal, and by $\pi$ bonding due to the back-donation of electrons from the filled $t_{2g}$ band of the transition metal to the unfilled $2\pi^{*}$ (LUMO) orbital of CO. Detailed study of the ``CO/Pt(111)" puzzle in the last few decades
	(as referenced below) suggests the following picture: The semilocal functionals like LDA and PBE yield incorrect energy levels of the highest occupied molecular orbital (HOMO) and lowest unoccupied molecular orbital (LUMO) of the adsorbed CO relative to the transition-metal Fermi level, and these incorrect levels are responsible for the incorrect prediction that the hollow site is the preferred or stable adsorption site. Self-interaction error (SIE) present in the popular local or semilocal exchange-correlation density functionals is to be blamed for this anomaly, placing the $2\pi^{*}$ LUMO too low relative to the metallic Fermi level. This enhances the back-donation of electronic charge from the metal to the oxygen side of the molecule. This strong delocalization of the electronic charge is the reason for the nonphysical adsorption of CO on the fcc hollow site of (111) surfaces of Cu, Rh, and Pt. This overestimation of the back-donation of the electronic charge also accounts for the too large adsorption energies from the semilocal functionals.
%	and the underestimation of the HOMO-LUMO gap of the adsorbed CO on the metal surface.
	
	Many methods and techniques have been prescribed over the past few decades to correct this back-donation problem due to the SIE. Among those the most computationally economic one is the selfconsistent ``DFT+U" method suggested by Kresse et al.\cite{CO-Kresse-DFTU} for the ``CO/Pt (111)" problem, where the empirical parameter U=0.75 eV is applied to achieve the correct top site adsorption position by adjusting the HOMO-LUMO gap of CO upward to the experimental transition energy. This method (combined with the PW91 GGA functional) was also applied successfully for Cu (111) surfaces by Gajdo\'{s} and Hafner\cite{GAJDOS2005DFTU}. Mason et al.\cite{Rappe-Mason-PRB} reported another empirical method to solve this back-donation problem for the CO adsorption on many metallic surfaces. In their method, CO singlet-triplet excitation energies from high-level coupled-cluster and configuration-integration method are used to obtain a linear relationship between the CO adsorption energy and the CO singlet-triplet splitting energy. These authors adjusted their C and O pseudopotential radii to reproduce the gas-phase coupled-cluster singlet-triplet splitting and  HOMO-LUMO energies. For $1/4$ monolayer of CO on Pt(111), the correction of Mason et al. changes\cite{DCMS} the dipole moment of an adsorbed molecule 
	from 0.026 to 0.089 $e\SI{}{\angstrom}$ (top site), and from -0.137 to -0.079 $e\SI{}{\angstrom}$ (fcc hollow site). 
	
	A fraction of exact exchange in the hybrid functionals tends to reduce the SIE in most cases, at moderate to very high computational cost compared to the LDA, GGAs and meta-GGAs. The successful hybrids PBE0\cite{PBE01,PBE02,PBE03}, B3LYP\cite{B3LYP,SDCF} and HSE\cite{hse03,hse06} can give more accurate descriptions of the adsorption energies for some metals but not all, and correct the HOMO-LUMO gap of the adsorbed CO. Wang et al.\cite{WangJACS2007}, found that PBE0 solved or nearly solved the ``CO/Pt(111) puzzle". Gil et al.\cite{GilSUSCI2003} had earlier found similar results from the B3LYP hybrid functional for the CO/Pt (111) problem, within a cluster model for Pt. These studies and those of Stroppa et al.\cite{STROPPAPRB2007} show that hybrid functionals create a near degeneracy between the top and hollow sites for CO/Pt(111). However, as pointed out by Stroppa and Kresse\cite{kresse-shortcoming}, the hybrid functionals often worsen the properties of bulk metals.
	
	Weak van der Waals (vdW) or dispersion interactions are largely missing in PBE, but intermediate-range vdW is included in PBEsol and SCAN, and long-range vdW in SCAN+rVV10. These interactions are not negligible for CO adsorption on metallic surfaces. For example, a study by Lazi\'{c} et al.\cite{Brako_PRB10} found that adding a nonlocal dispersion correction to PBE and revPBE tends to stabilize the top configuration as the preferred adsorption site. 
	
	In some recent studies\cite{BEEFPRB2012,mBEEFJCP2014} it has been reported that statistically-fitted GGA and meta-GGA functionals can reduce the error of CO adsorption energy 
	on transition metal surfaces. These functionals show improved chemisorption energies for adsorption of CO on a metallic surface, while underestimating surface energies compared to experiment. While the performance of the mBEEF meta-GGA for CO adsorption energies is similar to that of RPBE, both fail to yield the correct adsorption site for CO/Pt (111). However, mBEEF provides its own Bayesian error estimate.
	
	The fifth and highest rung of the ladder comprises RPA-like approximations that employ the unoccupied as well as the occupied orbitals. Ren et al.\cite{Ren_PRB} and Schimka et al.\cite{schimka2010accurate} found that the random phase approximation (applied to PBE orbitals and orbital energies) correctly described both CO adsorption energies and site preferences, while their tested semilocal functionals did not. A possible interpretation, consistent with similar results for the energy barriers to chemical reactions\cite{XRS}, is that RPA is relatively immune to density-driven error. In fact, Ren et al. found that the RPA adsorption energies for CO on Cu(111) are insensitive to the functional used to provide the orbital inputs. 
	
	In this work we aim to discuss this CO adsorption problem using the new meta-GGA SCAN. In Section II we discuss the details of the computational method used in this work. Section III summarizes our results. Section IV shows that the site-preference can be corrected and the adsorption energy can be improved for CO on Pt(111) by applying PBE not to its selfconsistent density but to a PBE+U density. The approximate self-interaction correction +U, applied to the atomic p orbitals of C and O, affects the total energy strongly for the adsorbed system, weakly for the separated CO molecule, and not at all for the separated Pt surface. This calculation strongly supports the idea that the important error in the PBE and SCAN descriptions of adsorption is driven by a charge-transfer error in the adsorbed system. In Section V we present our conclusions.

	\vspace*{0.5cm}
	\section{\label{sec:compdetail}COMPUTATIONAL DETAILS} 
	\vspace*{0.5cm}
	
	The results presented in this work are obtained from periodic density functional calculations performed with the Vienna Ab Initio Simulation (VASP)\cite{hafner2008ab}package. The projector-augmented-wave (PAW)\cite{blochl1994projector} method is used to describe the electron-ion interaction, with a plane-wave cut-off of 600 eV. Careful observation and previous studies revealed that a moderate to high cut-off energy can be used to study CO adsorption on transition metal surfaces. We used a PBE-PAW pseudo-potential for GGA and meta-GGA calculations, and an LDA PAW pseudo-potential for LDA calculations. The substrates in this work are modeled using 6 layers of metal with a 10 $\SI{}{\angstrom}$ vacuum region on top of $2\times2$ fcc surfaces of metals. Brillion-Zone sampling is done with a $12\times12\times1$ $\Gamma$-centered k mesh for the metallic surface slabs. To calculate the energetics of the CO molecule, we used a $15 \times 16 \times 17 \SI{}{\angstrom}$ box where the CO molecule is centered in the box. The top site is modeled by placing the C atom of CO directly on top of the metal atom, whereas the fcc and hcp hollow sites are modeled so that there is no metal atom directly under the C atom of CO in the second and first layer respectively. For all metallic surfaces, we chose a $1/4$ monolayer for computational efficiency and consistency. Table I of Lazi\'{c} et al.\cite{CO-Pt-Puzzle-4} shows that, for the four functionals tested there, the magnitude of the change in the adsorption energy of CO on the top site Pt(111) from $1/12$ to $1/4$ monolayer coverage is less than 0.025 eV and thus negligible for the purpose of our study. 

	We define the calculated CO adsorption energies as
			\vspace*{0.5cm}	
	\begin{equation}	
	E_{ads} =   E_{CO/M(111)}- (E_{CO} + E_{M(111)}),
	\label{eq:ads-1}
	\vspace*{.5cm}
	\end{equation}
	where $E_{CO/M(111)}$ is the total energy of the M (111) (M=Pd, Rh, Pt, Cu, Au, Ag) slab model with the adsorbed CO molecule, $E_{CO}$ is the energy of an isolated CO molecule, and $E_{M(111)}$ is the energy of the optimized clean $M(111)$ surface.

	\vspace*{0.5cm}	
	\section{\label{sec:result}RESULTS \& DISCUSSIONS} 
	\vspace*{0.5cm}
	
\begin{figure*}[!htb]
			\includegraphics[width=0.35\textwidth]{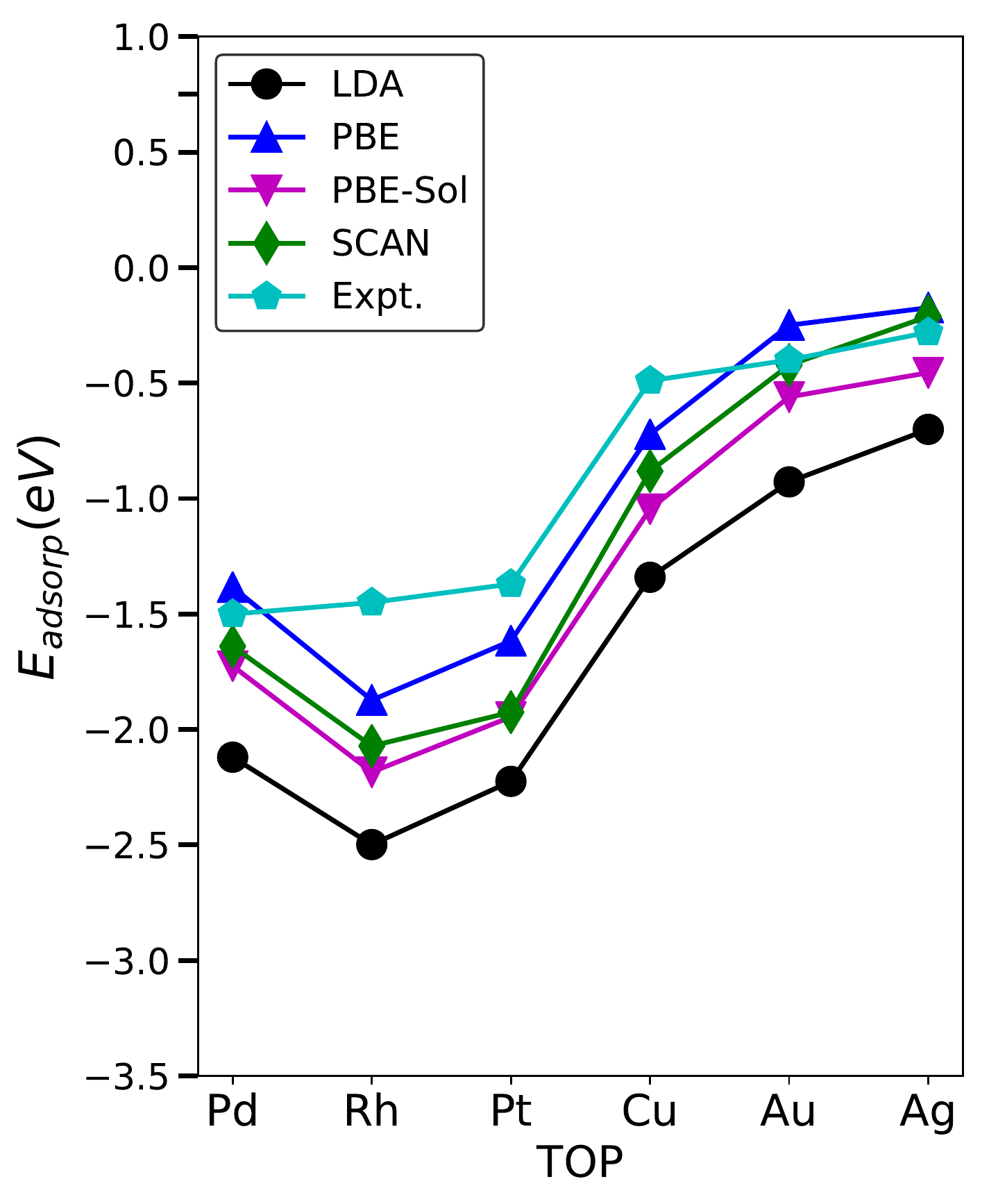}
\caption{Adsorption energy (in eV) of CO at the top site on different d -metals calculated using different functionals. Experimental adsorption energies are for the top site, except for Pd (111). For tables of calculated adsorption energies on various sites, see the Supplementary Information. For experimental adsorption energies and calculated site dependences, see Table \ref{tab:adsorp}.   
			}
			\label{fig:site-adsorp}
		\end{figure*}

	Table \ref{tab:adsorp}, Fig.\ref{fig:site-adsorp}, and the Supplementary Information (SI) give a complete compilation of calculated adsorption energies and related quantities for a CO molecule on several transition metal surfaces, including LDA, PBE, PBEsol, SCAN values. Table \ref{tab:adsorp} and Fig.\ref{fig:site-adsorp} also include the adsorption energy for the experimentally observed and thus most stable site. Figure \ref{fig:site-adsorp} shows that all functionals capture the correct chemical trends: Stronger binding on Pd, Rh, and Pt. where the d-bands of the metal have more energy overlap with the frontier orbitals of the molecule, and weaker binding on the noble metals Cu, Au, and Ag. But there are clear discrepancies between the calculated values of adsorption energies and the experimental data. The adsorption energy magnitudes are overestimated by all the functionals tested in this work (with the exception of PBE for CO on Au and Ag). This overestimation is greatest for LDA and least for PBE. SCAN values fall between PBE and PBEsol. 
%	Using the theoretically most-bound site, SCAN+rVV10 overestimates the adsorption energies by 54 \% for Pd (111), 28 \% for Rh (111) and 71 \% for Ag (111). 
	SCAN predicts the correct adsorption site for Pd(111) and 
	Rh(111), and comes close to predicting the correct site in
	Ag(111) and Pt(111). The difference in adsorption energies between the top and the fcc site for CO/Pt (111) from SCAN is 0.02 eV, better than the 0.15 eV from PBE. The functionals used in this work predict the wrong adsorption site for Cu (111), Au (111) and Pt (111).  The wrong adsorption site prediction by PBE, especially for ``CO/Pt(111)", is a long-standing challenge\cite{CO-Pt-Puzzle-1}. We do not see any changes in the PBE prediction from previously published results\cite{sun2011improved,Brako_PRB10,BEEFPRB2012}.
	
	The bond-length of the CO molecule and the distance between the surface metal atom and the C atom can be seen in Table S2 of SI. Our reported results from SCAN are in good agreement with LEED experimental data\cite{CO-Distance-LEED}. The experimentally observed C-O bond length ($d_{CO}$) of $1.15 \pm 0.05 \SI{}{\angstrom}$ for CO/Pt (111) is well reproduced by SCAN for all three sites. The $d_{Pt-CO}$ distance for the top site calculated from SCAN is within the experimental accuracy ($1.85 \pm 0.10 \SI{}{\angstrom}$) \citep{CO_PT_Distance,CO-Distance-LEED}. SCAN predicts a more accurate distance than PBE. For the other two (higher-coordinated) adsorption sites (fcc and hcp), the $d_{Pt-CO}$ distances are comparable for both SCAN and PBE. SCAN is more accurate than the other functionals for the metal-to-molecule binding distances in the other systems. 
	
\FloatBarrier
\vspace*{-15cm}
%    \FloatBarrier

	We have plotted the orbital-decomposed density of states (DOS) of the adsorbed systems using PBE in Figs, S2-S4 of the SI.

\vspace*{-3.5cm}
%    \FloatBarrier

	\begin{table*}[!htb]	
		\makebox[1 \textwidth][c]{       %centering table
			\resizebox{1.0 \textwidth}{!}{
				\begin{tabular}{l*{7}c}
						\hline
					\toprule
					 &  & 	&  &   &   & 	&  \\
					
					Methods & 	& \multicolumn{1}{c}{Pd} & \multicolumn{1}{c}{Rh}
					& \multicolumn{1}{c}{Pt}  & \multicolumn{1}{c}{Cu}  & \multicolumn{1}{c}{Au} 
					& \multicolumn{1}{c}{Ag} \\
					 &  & 	&  &   &   & 	&  \\
					\hline
					\midrule \addlinespace  \\
					%	\cmidrule(lr){14-15} \cmidrule(lr){14-15}
%					
%					& $\Delta E_{TOP-FCC}$  \hspace{0.3cm} & $\Delta E_{TOP-FCC}$   \hspace{0.3cm}  & $\Delta E_{TOP-FCC}$  \hspace{0.3cm}  & $\Delta E_{TOP-FCC}$  \hspace{0.3cm}
%					& $\Delta E_{TOP-FCC}$  \hspace{0.3cm} 
%					& $\Delta E_{TOP-FCC}$    \\
					% 		\cmidrule{3-6} \cmidrule{7-10} \cmidrule{11-14} \cmidrule{15-15}
					\midrule \addlinespace \\
					LDA   	& $ E_{TOP}$ & -2.12 \hspace{0.3cm} & -2.49 \hspace{0.3cm} & -2.22 \hspace{0.3cm}& -1.34 \hspace{0.3cm}& -0.93\hspace{0.3cm}& -0.70  ~\vspace*{0.5cm} \\
					
						& $ \Delta E_{TOP-FCC}$	& 0.78 \hspace{0.3cm} & 0.23 \hspace{0.3cm} & 0.35 \hspace{0.3cm}& 0.33 \hspace{0.3cm}& 0.23\hspace{0.3cm}& 0.18  ~\vspace*{0.5cm} \\

					PBEsol 		& $ E_{TOP}$	& -1.72 \hspace{0.3cm} & -2.18 \hspace{0.3cm} & -1.94 \hspace{0.3cm}& -1.05 \hspace{0.3cm}& -0.56 \hspace{0.3cm}& -0.45 ~\vspace*{0.5cm} \\
					
					& $ \Delta E_{TOP-FCC}$ & 0.71 \hspace{0.3cm} & 0.09 \hspace{0.3cm} & 0.27 \hspace{0.3cm}& 0.04 \hspace{0.3cm}& 0.15 \hspace{0.3cm}&  0.07 ~\vspace*{0.5cm} \\
					
					PBE 			& $ E_{TOP}$	& -1.38 \hspace{0.3cm} & -1.87 \hspace{0.3cm} & -1.61 \hspace{0.3cm}& -0.72 \hspace{0.3cm}& -0.25 \hspace{0.3cm}& -0.17  ~\vspace*{0.5cm} \\
					
						& $ \Delta E_{TOP-FCC}$ & 0.60 \hspace{0.3cm} & -0.03 \hspace{0.3cm} & 0.15 \hspace{0.3cm}& 0.07 \hspace{0.3cm}& 0.01 \hspace{0.3cm}& -0.05  ~\vspace*{0.5cm} \\

					SCAN 		& $ E_{TOP}$		& -1.64 \hspace{0.3cm} & -2.07 \hspace{0.3cm} & -1.92 \hspace{0.3cm}& -0.88 \hspace{0.3cm}& -0.42 \hspace{0.3cm}&  -0.21 ~\vspace*{0.5cm} \\
					
					& $ \Delta E_{TOP-FCC}$ & 0.60 \hspace{0.3cm} & -0.12 \hspace{0.3cm} & 0.02 \hspace{0.3cm}& 0.13 \hspace{0.3cm}& 0.03 \hspace{0.3cm}&  0.00 ~\vspace*{0.5cm} \\
				
					Expt. 	&		&  $-1.48\pm 0.09^{\cite{COPDEXPT}}$ (F)  \hspace{0.3cm} 
					& $-1.45(T)^{\cite{CORHEXPT}}$ \hspace{0.3cm}
					& $-1.37(T)^{\cite{COPTEXPT}}$  \hspace{0.3cm} 
					& $-0.50 \pm 0.05(T)^{\cite{COCUEXPT}}$ \hspace{0.3cm}
					& $-0.40(T)^{\cite{COAUExpt}}$    \hspace{0.3cm}
					& $-0.28(T)^{\cite{MCELHINEY1976617}}$  \\
					
				&  &  \hspace{0.3cm} &  \hspace{0.3cm} & \hspace{0.3cm}&  \hspace{0.3cm}&  \hspace{0.3cm}&  ~\vspace*{0.5cm} \\
					\hline
					\bottomrule
				\end{tabular}
			}
		}
		\caption{The first row for each functional shows the top-site adsorption energy (in eV) for CO on the (111) surface. The second row for each functional shows the calculated difference of top and fcc-hollow adsorption energies (in eV) for CO on (111) surfaces of different \textit{d}-metals. The letter (T= top, F= fcc, H=hcp) in the parentheses of the last row indicates the stable adsorption site from experiment. For a correct site prediction, the energy difference $\Delta E_{TOP-FCC}$ should be negative for the T cases and positive for the F case. The numbers in the last row are experimental adsorption energies. }
		\label{tab:adsorp}
		\begin{footnotesize}

		\end{footnotesize}
	\end{table*}

  \FloatBarrier
\vspace*{0.5cm}
\section{\label{sec:corerction}CORRECTING THE DENSITY FOR CO on Pt(111): PBE@PBE+U} 
\vspace*{0.5cm}	
	
	We have performed DFT+U calculations on PBE structures in order to investigate if the localized self-interaction correction of the DFT+U method \cite{Cococcioni_PRB} can be used to remove the density-driven error of a self-consistent PBE calculation. To do that, we applied a series of increasing U values. Table II shows total energies for the adsorbed system, the clean surface, and the free molecule, for the PBE functional. These energies and thus the adsorption energies are calculated in a non-self-consistent way, using the self-consistent PBE+U density as shown in Eq 2. It is important to note that U has only been applied on the atomic p- orbitals in the free and adsorbed CO molecule. The adsorption energies are calculated in the following way-
	
		\begin{align}	
	E_{ads}^{non-SCF} =&   E^{non-SCF-PBE@PBE+U}_{CO/Pt(111)}- \\
	+ & (E^{non-SCF-PBE@PBE+U}_{CO} \nonumber \\ 
	+& E^{non-SCF-PBE@PBE+U}_{Pt(111)}) \nonumber \\	
	\nonumber \\	
	E_{ads}^{SCF} =&   E^{SCF-PBE}_{CO/Pt(111)}- (E^{SCF-PBE}_{CO} + E^{SCF-PBE}_{Pt(111)}).
	\vspace*{.5cm}
	\label{eq:scan-pbe+u}
	\end{align}
	
	 In Table S3 of the Supplementary Materials, we similarly apply the SCAN functional to the PBE+U density on the SCAN structures. 
	
	Table II shows that the +U correction to the PBE density has no effect at all on the total energy of the clean surface, and negligibly changes the PBE total energy of the free molecule (by less than 0.015 eV), but it significantly raises the total energy of the adsorbed system (by as much as 0.5 eV). Increasing U from 0 to +0.75 eV correctly stabilizes the top-site adsorption, and brings the adsorption energy on that site into agreement with experiment. Further increases in U up to +2.00 eV have little further effect on the adsorption energy. This shows that U has an important and correct effect on the electron density and total energy only when the separate systems are brought together and allowed to exchange electrons. We conclude that the errors of the SCAN functional for this system are largely density-driven and arise from the incorrect charge transfers that plague all semilocal approximations to the density functional. 

%	Table II. Total energies and adsorption energies of an adsorbed CO molecule on Pt (111) surfaces for three different configurations, computed non -selfconsistently with a self-consistent PBE+U density with nonzero U values. The energies for a non-selfconsistent calculation with U = 0 and a self-consistent PBE calculation agree very well, as expected. These energies are determined from the total energies of the combined CO/Pt (111) system, clean Pt(111) surface, and free CO molecule, using Eqs. 2 and 3. 
%	

	\begin{table}[]
		\begin{tabular}{lllllll}
			\hline
				&	&				& 		&  		&    &   \\
			&	&	& CO/Pt(111)    & Pt(111) & CO  & $E_{Ads}$ \\
				&	&				& 		&  		&    &   \\
			
			\hline
			&	&	&    			&  		  &   	&  			\\
			&  non-SCF-PBE	&	&    			&  		  &   	&  		\\
			& @PBE+U	&	&    			&  		  &   	&  			\\
			&  (U = 0.00) & TOP	& -157.70 & -141.46 & -14.63 & -1.61 \\
			&	& FCC	&  -157.85 & -141.46 & -14.63 & -1.76 \\
			&	& HCP	& -157.83 &-141.46 & -14.63  & -1.74 \\
			
			& non-SCF-PBE	&	&    			&  		  &   	&  			\\
			& @PBE+U	&	&    			&  		  &   	&  			\\
			& (U = 0.40)	& TOP	&  -157.49 & -141.46 & -14.63& -1.40 \\
			&	& FCC	&  -157.42 & -141.46 & -14.63 & -1.34 \\
			&	& HCP	& -157.42 & -141.46 & -14.63 & -1.33 \\
			
			& non-SCF-PBE	&	&    			&  		  &   	&  			\\
			& @PBE+U	&	&    			&  		  &   	&  			\\
			& (U = 0.75)	& TOP	& -157.45 & -141.46 & -14.63 & -1.36   \\
			&	& FCC	& -157.37 & -141.46 & -14.63 & -1.28 \\
			&	& HCP	& -157.37  & -141.46 & -14.63 & -1.28 \\
			
			& non-SCF-PBE	&	&    			&  		  &   	&  			\\
			& @PBE+U	&	&    			&  		  &   	&  			\\
			& (U = 1.00)  & TOP	& -157.44 & -141.46 & -14.63 & -1.35  \\
			&	& FCC	&  -157.35 & -141.46 & -14.63 & -1.26 \\
			&	& HCP	& -157.35 & -141.46 & -14.63 & -1.26   \\
			
			& non-SCF-PBE	&	&    			&  		  &   	&  			\\
			& @PBE+U	&	&    			&  		  &   	&  			\\
			& (U = 1.25)  & TOP	& -157.44 & -141.46 & -14.64 & -1.34  \\
			&	& FCC	& -157.35 & -141.46 & -14.64 & -1.26   \\
			&	& HCP	& -157.35 & -141.46 & -14.64 & -1.26    \\
			
			& non-SCF-PBE	&	&    			&  		  &   	&  			\\
			& @PBE+U	&	&    			&  		  &   	&  			\\
			& (U = 1.50)  & TOP	& -157.43 & -141.46 & -14.64 & -1.33  \\
			&	& FCC	& -157.33 & -141.46& -14.64 & -1.23  \\
			&	& HCP	& -157.34 & -141.46 & -14.64 & -1.24  \\
			
			& non-SCF-PBE	&	&    			&  		  &   	&  			\\
			& @PBE+U	&	&    			&  		  &   	&  			\\
			& (U = 1.75)  & TOP	& -157.42 & -141.46  & -14.64 & -1.32   \\
			&	& FCC	& -157.33 & -141.46  & -14.64 & -1.23  \\
			&	& HCP	& -157.29 & -141.46  &-14.64 & -1.19 \\
			
			& non-SCF-PBE	&	&    			&  		  &   	&  			\\
			& @PBE+U	&	&    			&  		  &   	&  			\\
			& (U = 2.00)  & TOP	& -157.42 & -141.46 & -14.64 & -1.31   \\
			&	& FCC	& -157.32 & -141.46 & -14.64 & -1.22   \\
			&	& HCP	& -157.27 & -141.46 & -14.64 & -1.17  \\
			
			&  SCF-PBE	&	&    			&  		  &   	&  			\\
			&  @PBE	&	&    			&  		  &   	&  			\\
			&   & TOP	& -157.71 & -141.46 & -14.63 & -1.61 \\
			&	& FCC	& -157.85 & -141.46 & -14.63 & -1.76 \\
			&	& HCP	& -157.83 & -141.46 & -14.63 & -1.74 \\
			&	&		&    			&  		 	 &   		&  			\\

			\hline
		\end{tabular}
		\caption{Total energies and adsorption energies (in eV) of an adsorbed CO molecule on Pt (111) surfaces for three different configurations, computed non -selfconsistently with a self-consistent PBE+U density with nonzero U values. The energies for a non-selfconsistent calculation with U = 0 and a self-consistent PBE calculation agree very well, as expected. These energies are determined from the total energies of the combined CO/Pt (111) system, clean Pt(111) surface, and free CO molecule, using Eqs. 2 and 3. }
		\label{tab:pbe-energies}
	\end{table}

The projected densities of states for C and O (see Fig. S2 of SI) on the surface show very small changes with U. The gross features of these plots are that the C atom in adsorbed CO has about two p electrons and four p holes, while the O atom in adsorbed CO has about four p electrons and two p holes, regardless of U. Thus the changes in the orbital energies due to U are small and subtle, presumably due to metallic screening. We have resisted the temptation to quantify the density change due to U, except insofar as it changes the total energies and their differences, since all density partitioning schemes are somewhat arbitrary. A change in dipole moment is not arbitrary, but there is no unique way to separate it into a contribution from charge transfer within the molecule and another from charge transfer between the molecule and the surface.

  \FloatBarrier
	\vspace*{0.5cm}
	\section{\label{sec:conclu}CONCLUSION} 
	\vspace*{0.5cm}
The adsorptions we discuss are mostly overbound by the PBE GGA\cite{PBE}, and are overbound even more by the SCAN meta-GGA\cite{SCAN}, in self-consistent calculations. Self-consistent PBE normally overestimates covalent bond strengths (while underestimating vdW bond strengths), and SCAN is normally much more accurate than PBE for all kinds of bonds in both molecules and condensed matter (as detailed in section I). Any deviation from the self-consistent density of the bonded system for a given functional will by the variational principle reduce the computed binding, as needed for CO on a transition-metal surface. In most practical calculations the error in the DFT energy is primarily functional-driven. However, in a few special situations (e.g., when a good semilocal functional is applied to a system with a small or zero HOMO-LUMO gap and possible charge transfer)\cite{KeironJCP2014} the error is primarily density driven. Here we might say that the zero energy gap of the metal allows a large density response to errors in the PBE or SCAN exchange-correlation potential of the adsorbed molecule. It is known that the density-driven error can be sensitive to the specific exchange-correlation potential, and can be cured by using a more accurate density then the self-consistent density of DFT. Often, the Hartree-Fock (HF) density\cite{KeironPRL2013} serves this purpose for molecules and anions. Hartree-Fock theory is self-interaction-free and greatly reduces charge-transfer errors. 
	
When this article was first submitted, it ended with this statement: ``We would have liked to apply SCAN to a Hartree-Fock or hybrid-functional density for CO on transition-metal surfaces, to check if this procedure yields the correct binding energies and adsorption sites. But the computational cost of such a calculation is high. We are looking for alternatives, e.g., the self-interaction correction\citep{PZSIC,PedersonSIC} or the local orbital scaling correction.\cite{WangSIC}'' Both referees suggested that we try the PBE+U method for the electron density. Without an expensive first-principles determination of U \cite{Cococcioni_PRB}, this is an empirical but computationally efficient method. The results of its implementation in section IV strongly support the contention of this paper, that the errors of SCAN (and other semilocal functionals) for CO adsorption on transition metal surfaces are largely density driven. A fully-satisfactory solution to the CO/Pt(111) puzzle might require a good nonempirical self-interaction correction to the semilocal functionals.

For a given functional like PBE, and for a given energy difference like the adsorption energy or the difference of adsorption energy among competing adsorption sites, there is a density error. If correcting that error makes a big improvement in that energy difference for that functional, we can say unequivocally that the energy-difference error is density driven. Density-driven error depends upon the functional and upon the energy difference. For a particular energy difference, some functionals (like RPA) may be less sensitive to density error than other functionals (like PBE). This fact could explain why RPA@PBE is much more correct \cite{schimka2010accurate} for the site preference and adsorption energy of CO on Pt(111) than is PBE@PBE, even though the density error is the same for both.

\vspace*{0.5cm}
\section{\label{sec:ackno}ACKNOWLEDGMENTS} 
\vspace*{0.5cm}
The work of AP and JPP was supported by the U.S. National Science Foundation under Grant No. DMR-1607868 (CMMT – Division of Materials Research, with a contribution from CTMC – Division of Chemistry). The work of HP and JS was supported by the U.S. Department of Energy, Office of Basic Energy Sciences, through the Energy Frontier Research Center “Center for Complex Materials from first Principles”, under grant DE-SC0012575.

	\FloatBarrier	
	\bibliography{co-vdw-2017}{}
	\bibliographystyle{unsrtnat}
	
\end{document}

\title{Supplementary Information for\\ ``Re-thinking CO adsorption on transition-metal surfaces: Density-driven error?"}

	\author{Abhirup Patra}
\affiliation{Department of Physics, Temple University, Philadelphia, PA 19122}
\affiliation{School of Materials Science and Engineering, Georgia Institute of Technology, Atlanta, GA 30308}
\author{Haowei Peng}
\affiliation{Department of Physics, Temple University, Philadelphia, PA 19122}
\author{Jianwei Sun}
\affiliation{Department of Physics and Engineering Physics, Tulane University, New Orleans, LA 70118}
\author{John P. Perdew}
\affiliation{Departments of Physics and Chemistry, Temple University, Philadelphia, PA 19122}

%\date{}
\maketitle

%\section{Adsorption energies}
\label{sec:adsorption}
\renewcommand\thetable{S1}
\begin{table}[H]	
	\makebox[1 \textwidth][c]{       %centering table
		\resizebox{1.0 \textwidth}{!}{
			\begin{tabular}{l*{18}c}
				\hline
				\toprule
				 & \multicolumn{3}{c}{} & \multicolumn{3}{c}{}
				& \multicolumn{3}{c}{}  & \multicolumn{3}{c}{}  & \multicolumn{3}{c}{} 
				& \multicolumn{3}{c}{} \\
				
				Methods & \multicolumn{3}{c}{Pd} & \multicolumn{3}{c}{Rh}
				& \multicolumn{3}{c}{Pt}  & \multicolumn{3}{c}{Cu}  & \multicolumn{3}{c}{Au} 
				& \multicolumn{3}{c}{Ag} \\
				
				 & \multicolumn{3}{c}{} & \multicolumn{3}{c}{}
			& \multicolumn{3}{c}{}  & \multicolumn{3}{c}{}  & \multicolumn{3}{c}{} 
			& \multicolumn{3}{c}{} \\
				\hline
				\midrule \addlinespace  \\
				%	\cmidrule(lr){14-15} \cmidrule(lr){14-15}
				
				& TOP & FCC & HCP \hspace{0.3cm}& TOP & FCC & HCP  \hspace{0.3cm}& TOP & FCC & HCP \hspace{0.3cm} & TOP & FCC & HCP \hspace{0.3cm}& TOP & FCC & HCP\hspace{0.3cm} & TOP & FCC & HCP  \\
				\cmidrule{3-6} \cmidrule{7-10} \cmidrule{11-14} \cmidrule{15-15}
				\midrule \addlinespace \\
				LDA 		& -2.12 & -2.90 & \textbf{-2.93 }\hspace{0.3cm} & -2.49 & -2.72 & \textbf{-2.85} \hspace{0.3cm} & -2.22 & \textbf{-2.57} & -2.56 \hspace{0.3cm}& -1.34 & -1.67 & \textbf{-1.64} \hspace{0.3cm}& -0.93 & \textbf{-1.16} & -1.08 \hspace{0.3cm}& -0.70 & -0.88 & \textbf{-1.08}   ~\vspace*{0.5cm} \\
				
				PBEsol 		& -1.72 & -2.43 & \textbf{-2.44} \hspace{0.3cm}& -2.18 & -2.27 & \textbf{-2.36} \hspace{0.3cm}& -1.94 & \textbf{-2.21} & -2.20 \hspace{0.3cm}& -1.05 & -1.09 & \textbf{-1.28} \hspace{0.3cm}& -0.56 & \textbf{-0.71} & -0.66 \hspace{0.3cm}& -0.45 & -0.52 & \textbf{-0.66}  ~\vspace*{0.5cm} \\

				PBE 		& -1.38 & -1.98 & \textbf{-1.99} \hspace{0.3cm} & -1.87 & -1.84 & \textbf{-1.93} \hspace{0.3cm} & -1.61 & \textbf{-1.76} & -1.74 \hspace{0.3cm}& -0.72 & \textbf{-0.79} & -0.77 \hspace{0.3cm}& -0.25 & \textbf{-0.26} & -0.21 \hspace{0.3cm}& \textbf{-0.17} & -0.12 & -0.11  ~\vspace*{0.5cm} \\

				SCAN 		& -1.64 & \textbf{-2.24} & -2.22 \hspace{0.3cm} & \textbf{-2.07} & -1.95 & -2.03 \hspace{0.3cm}& -1.92 & \textbf{-1.94} & -1.89 \hspace{0.3cm}& -0.88 & \textbf{-1.01} & -0.99 \hspace{0.3cm}& -0.42 & \textbf{-0.45} & -0.38 \hspace{0.3cm}& -0.21 & \textbf{-0.21} & -0.20  ~\vspace*{0.5cm} \\

				Expt. 		&  &  $-1.48\pm 0.09$\textsuperscript{\cite{COPDEXPT}} (F) & \hspace{0.3cm} & $-1.45(T)$\textsuperscript{\cite{CORHEXPT}} &       & \hspace{0.3cm}& $-1.37(T)$\textsuperscript{\cite{COPTEXPT}} &       & \hspace{0.3cm} & $-0.50 \pm 0.05(T)$\textsuperscript{\cite{COCUEXPT}} &       &       \hspace{0.3cm}& -0.40(T)\textsuperscript{\cite{COAUExpt}} &       &       \hspace{0.3cm}& $-0.28(T)$\textsuperscript{\cite{MCELHINEY1976617}} &       &   \\
	
				\hline
	
				\bottomrule
			\end{tabular}
		}
	}
	\caption{Adsorption energies (in eV) for CO 
		on (111) surfaces of different d-metals. Boldface numbers
		indicate the favored adsorption site for each metallic
		surface predicted by each functional. The letter (T=top,
		F=fcc, H=hcp) in the parentheses of the experimental 
		adsorption energy row indicates the stable adsorption 
		site found in experiments.}
	\label{tab:SI-adsorp}
	% 	\begin{footnotesize}
	% 			
	% 	\end{footnotesize}
\end{table}

\FloatBarrier
%\section{Binding distances}
\label{sec:binding-distance}
\renewcommand\thetable{S2}
\begin{table}[!h]
	\makebox[1 \textwidth][c]{       %centering table
		\resizebox{1.0 \textwidth}{!}{ 	
			\begin{tabular}{ll*{18}c}
				\toprule
				Methods & &\multicolumn{3}{c}{Pd} & \multicolumn{3}{c}{Rh}
				& \multicolumn{3}{c}{Pt}  & \multicolumn{3}{c}{Cu}  & \multicolumn{3}{c}{Au} 
				& \multicolumn{3}{c}{Ag} \\
				\hline
				\midrule \addlinespace  \\
				%	\cmidrule(lr){14-15} \cmidrule(lr){14-15}
				
				& & TOP & FCC & HCP \hspace{0.3cm}& TOP & FCC & HCP  \hspace{0.3cm}& TOP & FCC & HCP \hspace{0.3cm} & TOP & FCC & HCP \hspace{0.3cm}& TOP & FCC & HCP\hspace{0.3cm} & TOP & FCC & HCP  \\
				\cmidrule{3-6} \cmidrule{7-10} \cmidrule{11-14} \cmidrule{15-15}
				\midrule \addlinespace \\
				LDA 	& $d_{C-O}$	& 1.15 & 1.18 & 1.19 \hspace{0.3cm} & 1.15 & 1.18 & 1.19 \hspace{0.3cm} & 1.15 & 1.18 & 1.18 \hspace{0.3cm}& 1.15 & 1.17 &  1.17 \hspace{0.3cm}& 1.15 & 1.17 & 1.17 \hspace{0.3cm}& 1.15 & 1.17 & 1.17   ~\vspace*{0.5cm} \\
				
				& $d_{M-C}$ & 2.02 & 1.35 & 1.24 \hspace{0.3cm} & 2.00 & 1.42 & 1.30 \hspace{0.3cm} & 1.96 & 1.43 & 1.32 \hspace{0.3cm}& 1.98 & 1.43 & 1.36 \hspace{0.3cm}& 2.21 & 1.45 & 1.34 \hspace{0.3cm}& 1.97 & 1.57  & 1.29   ~\vspace*{0.5cm} \\

				PBEsol 		& $d_{C-O}$	& 1.15 & 1.19 & 1.19 \hspace{0.3cm} & 1.16 & 1.19 & 1.19 \hspace{0.3cm} & 1.15 & 1.19 & 1.18 \hspace{0.3cm}& 1.16 & 1.18 & 1.18\hspace{0.3cm}& 1.15 & 1.18 & 1.18 \hspace{0.3cm}& 1.15  & 1.17 & 1.17   ~\vspace*{0.5cm} \\
				
				& $d_{M-C}$ & 1.87 & 1.35 & 1.30 \hspace{0.3cm} & 1.97 & 1.47 & 1.45 \hspace{0.3cm} & 1.93 & 1.45 & 1.32 \hspace{0.3cm}& 1.87 & 1.54 & 1.40 \hspace{0.3cm}& 2.05 & 1.42 & 1.35 \hspace{0.3cm}& 2.10 & 1.58 & 1.52   ~\vspace*{0.5cm} \\

				PBE 	& $d_{C-O}$	& 1.16 & 1.19 & 1.19 \hspace{0.3cm} & 1.16 & 1.19 & 1.19 \hspace{0.3cm} & 1.16 & 1.19 & 1.19 \hspace{0.3cm}& 1.15 & 1.18 & 1.18 \hspace{0.3cm}& 1.15 & 1.18 & 1.18 \hspace{0.3cm}& 1.15 & 1.17 & 1.17   ~\vspace*{0.5cm} \\
				
				& $d_{M-C}$ & 1.87 & 1.40 & 1.29 \hspace{0.3cm} & 2.00 & 1.46 & 1.35 \hspace{0.3cm} & 2.00 & 1.46 & 1.33 \hspace{0.3cm}& 1.85 & 1.57 & 1.43 \hspace{0.3cm}& 2.06 & 1.43 & 1.36 \hspace{0.3cm}& 2.09 & 1.66 & 1.58   ~\vspace*{0.5cm} \\

				SCAN 	& $d_{C-O}$	& 1.15 & 1.18 & 1.18 \hspace{0.3cm} & 1.15 & 1.18 & 1.19 \hspace{0.3cm} & 1.14 & 1.18 & 1.19 \hspace{0.3cm}& 1.15 & 1.17 & 1.17 \hspace{0.3cm}& 1.14 & 1.17 & 1.17 \hspace{0.3cm}& 1.15 & 1.16 & 1.16    ~\vspace*{0.5cm} \\
				
				& $d_{M-C}$ & 1.85 & 1.42 & 1.29 \hspace{0.3cm} & 1.99 & 1.45 & 1.33 \hspace{0.3cm} & 1.94 & 1.47 & 1.32 \hspace{0.3cm}& 1.91 & 1.55 & 1.39 \hspace{0.3cm}& 2.04 & 1.41 & 1.32 \hspace{0.3cm}& 2.02 & 1.66 & 1.54  ~\vspace*{0.5cm} \\

				Expt. 		& $d_{C-O}$	& &  & $1.14 \pm .014$ \textsuperscript{\cite{GIEEL199890}} \hspace{0.3cm} & $1.20\pm 0.05$\textsuperscript{\cite{GIERER1997176}} &  & \hspace{0.3cm} & $1.15\pm 0.1$\textsuperscript{\cite{CO-Distance-LEED}} &  &  \hspace{0.3cm}&  &  &  \hspace{0.3cm}&  &  & \hspace{0.3cm}&  &  &    ~\vspace*{0.5cm} \\
				
				& $d_{M-C}$ & & & $1.27\pm 0.04$\textsuperscript{\cite{GIEEL199890}} \hspace{0.3cm} & $1.87 \pm 0.04$\textsuperscript{\cite{GIERER1997176}} &  &   \hspace{0.3cm} & $1.85 \pm 0.1$\textsuperscript{\cite{CO-Distance-LEED,CO_PT_Distance}}  &  &  \hspace{0.3cm}& $1.91\pm 0.01$\textsuperscript{\cite{Cu-CO-Distance}}  &  &  \hspace{0.3cm}& &  &  \hspace{0.3cm}&  &  &   ~\vspace*{0.5cm} \\
				\hline
				\bottomrule
			\end{tabular}
		}
	}
	\caption{Calculated values of C-O bond length (in $\SI{}{\angstrom}$) of adsorbed CO on metals. The metal-to-carbon ($d_{M-C}$) distances (also in $\SI{}{\angstrom}$) are also tabulated for different metals.}
	\label{tab:distance}
\end{table}

\vspace*{-1cm}
\FloatBarrier
\begin{figure*}[!ht]
	\renewcommand\thefigure{S1}
	\includegraphics[width=1.0\textwidth]{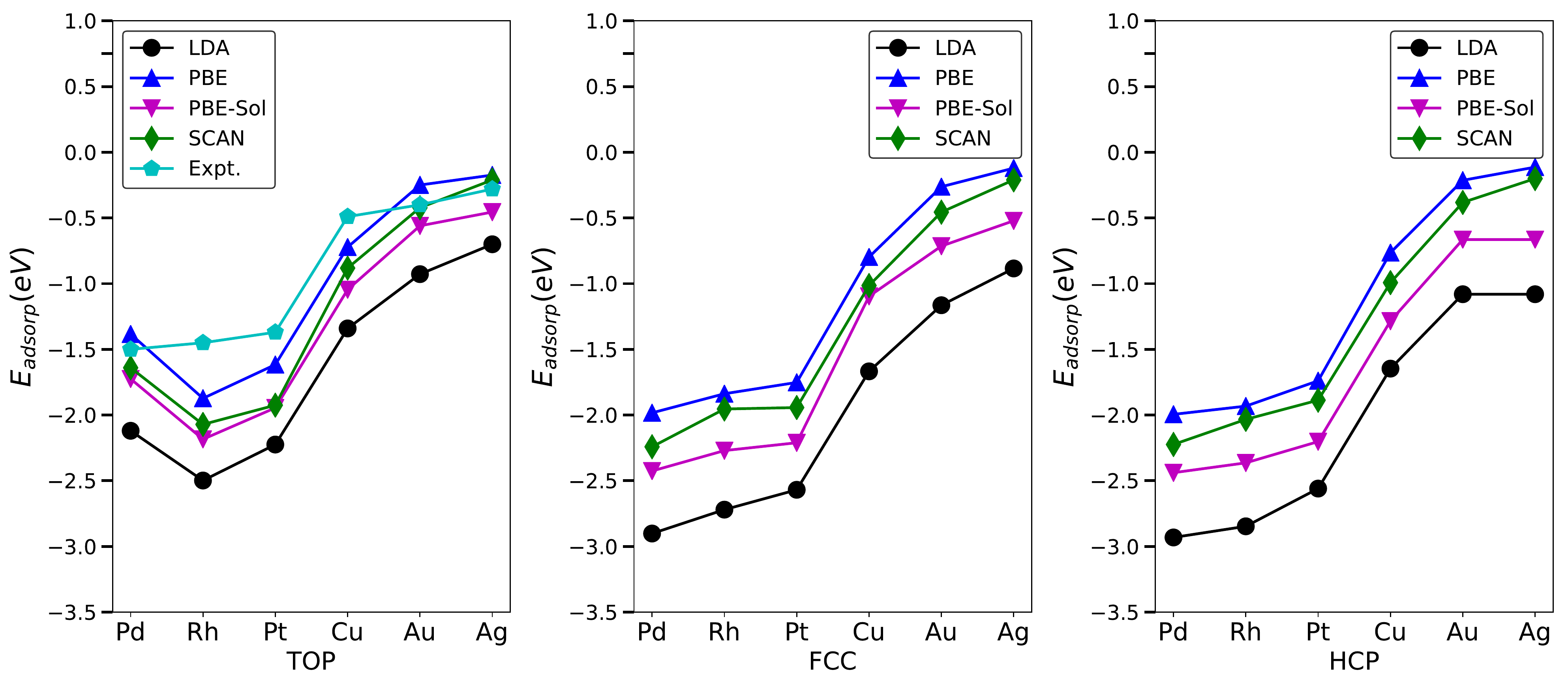}
	\caption{Adsorption energies (in eV) of CO on different \textit{d}-metals for different adsorption sites as tabulated in \ref{tab:SI-adsorp}. The experimental value in the leftmost panel is for the top site, except in Pd where it is for the FCC site.}
	\label{fig:site-adsorp}
\end{figure*}

\vspace*{2.0 cm}
We have performed DFT+U calculations on PBE and SCAN structures in order to investigate if the localized self-interaction correction of the DFT+U method can be used to remove the density-driven error from a selfconsistent PBE or SCAN calculation. To do that, we applied a series of increasing +U values to the adsorbed molecular orbital. Table 2 of the main text shows total energies for the adsorbed system, clean surface and free molecule for the PBE functional. These energies and thus the adsorption energies are calculated in a non-selfconsistent way from the selfconsistent PBE+U density. It is important to note that U has only been applied on the p-orbitals of the free and adsorbed CO molecule. The adsorption energies are calculated via Eqs. (2) and (3) of the main text.   

\vspace*{.35 cm}
From, TABLE S3 it can be seen that SCAN with the PBE density (U=0) already recovers the correct preference for top-site adsorption. Increasing U reduces SCAN overbinding, but some overbinding remains even at U=2 eV.

\FloatBarrier
\renewcommand\thetable{S3}
\begin{table}[]
	\begin{tabular}{lllllll}
		\hline
		&	&	&    			&  		  &   	&  			\\
		&	&	& CO/Pt(111)    & Pt(111) & CO  & $E_{Ads}$ \\
		&	&	&    			&  		  &   	&  			\\
		\hline
		& non-SCF-SCAN	&	&    			&  		  &   	&  			\\
		& @PBE+U	&	&    			&  		  &   	&  			\\
		& (U = 0.00)	& TOP	&	-1699.05& -1,680.52 & -16.61 & -1.92 \\
		&	& FCC	&		-1698.93 & -1,680.52 &  -16.61 & -1.80 \\
		&	& HCP	&		-1698.81 & -1,680.52 &  -16.61 & -1.58 \\
		
		& non-SCF-SCAN	&	&    			&  		  &   	&  			\\
		& @PBE+U	&	&    			&  		  &   	&  			\\
		& (U = 0.75)	& TOP	&-1699.06 & -1,680.52 & -16.62 & -1.92   \\
		&	& FCC	& -1698.83 & -1,680.52 & -16.62 & -1.70   \\
		&	& HCP	&-1698.68  & -1,680.52 & -16.62 & -1.55   \\
		
		& non-SCF-SCAN	&	&    			&  		  &   	&  			\\
		& @PBE+U	&	&    			&  		  &   	&  			\\
		& (U = 1.00)  & TOP	&-1699.04 & -1,680.52 & -16.62 & -1.91   \\
		&	& FCC	&-1698.80 & -1,680.52 & -16.62 & -1.67   \\
		&	& HCP	&-1698.68 & -1,680.52 & -16.62 & -1.54   \\
		
		& non-SCF-SCAN	&	&    			&  		  &   	&  			\\
		& @PBE+U	&	&    			&  		  &   	&  			\\
		& (U = 1.25)  & TOP	&-1699.01 & -1,680.52 & -16.62 & -1.88   \\
		&	& FCC	&-1698.74 & -1,680.52 & -16.62 & -1.60   \\
		&	& HCP	&-1698.67 & -1,680.52 & -16.62 & -1.54   \\
		
		& non-SCF-SCAN	&	&    			&  		  &   	&  			\\
		& @PBE+U	&	&    			&  		  &   	&  			\\
		& (U = 1.50)  & TOP	&-1698.91 & -1,680.52 & -16.61 & -1.77   \\
		&	& FCC	&-1698.69 & -1,680.52 & -16.61 & -1.57   \\
		&	& HCP	&-1698.67 & -1,680.52 & -16.61 & -1.53  \\
		
		& non-SCF-SCAN	&	&    			&  		  &   	&  			\\
		& @PBE+U	&	&    			&  		  &   	&  			\\
		& (U = 1.75)  & TOP	&-1698.86 & -1,680.52 & -16.61 & -1.73   \\
		&	& FCC	& -1698.66 & -1,680.52 & -16.61 & -1.53   \\
		&	& HCP	& -1698.64 & -1,680.52 & -16.61 & -1.53  \\
		
		& non-SCF-SCAN	&	&    			&  		  &   	&  			\\
		& @PBE+U	&	&    			&  		  &   	&  			\\
		& (U = 2.00)  & TOP	& -1698.75 & -1,680.52 & -16.61 & -1.62   \\
		&	& FCC	& -1698.62 & -1,680.52 & -16.61 & -1.49   \\
		&	& HCP	& -1698.61 & -1,680.52 & -16.61 & -1.48  \\
		
		&  SCF-SCAN	&	&    			&  		  &   	&  			\\
		&  @SCAN	&	&    			&  		  &   	&  			\\
		&	& TOP	& -1699.12 & -1680.57 & -16.62 & -1.92 \\
		&	& FCC	& -1699.13 & -1680.57 & -16.62 & -1.94 \\
		&	& HCP	& -1699.08 & -1680.57 & -16.62 & -1.89 \\
		&	&	&    			&  		  &   	&  			\\

		\hline
	\end{tabular}
	
	\caption{SCAN total energies (in eV) calculated non-self consistently using the PBE+U density for different values of U.}
	\label{tab:scan-energies}
\end{table}

\vspace{1 cm}

\FloatBarrier
\renewcommand\thefigure{S2}
\begin{figure*}[!h]
	
	\centering
	\includegraphics[width=0.35\textwidth]{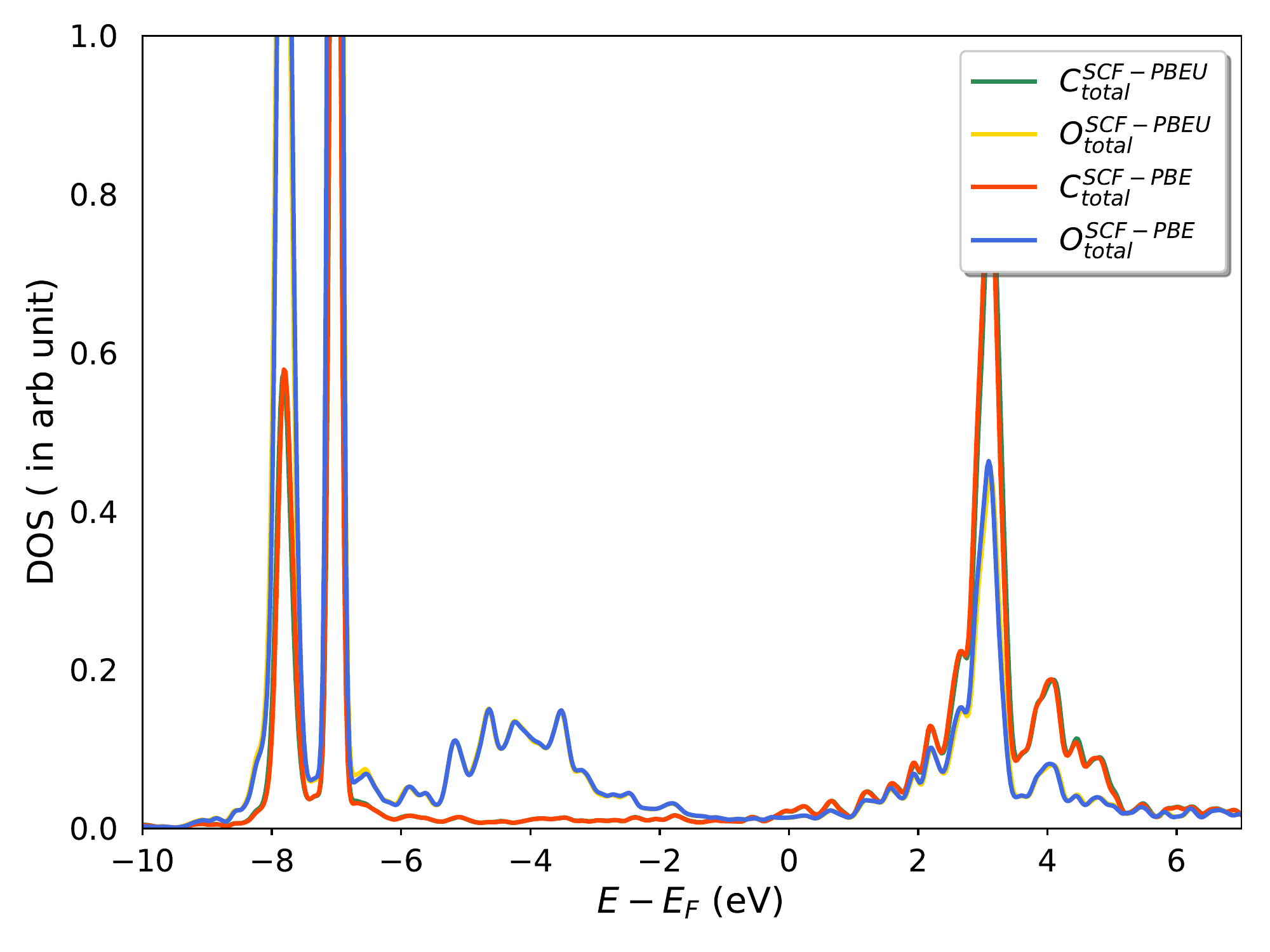}~
	\includegraphics[width=0.35\textwidth]{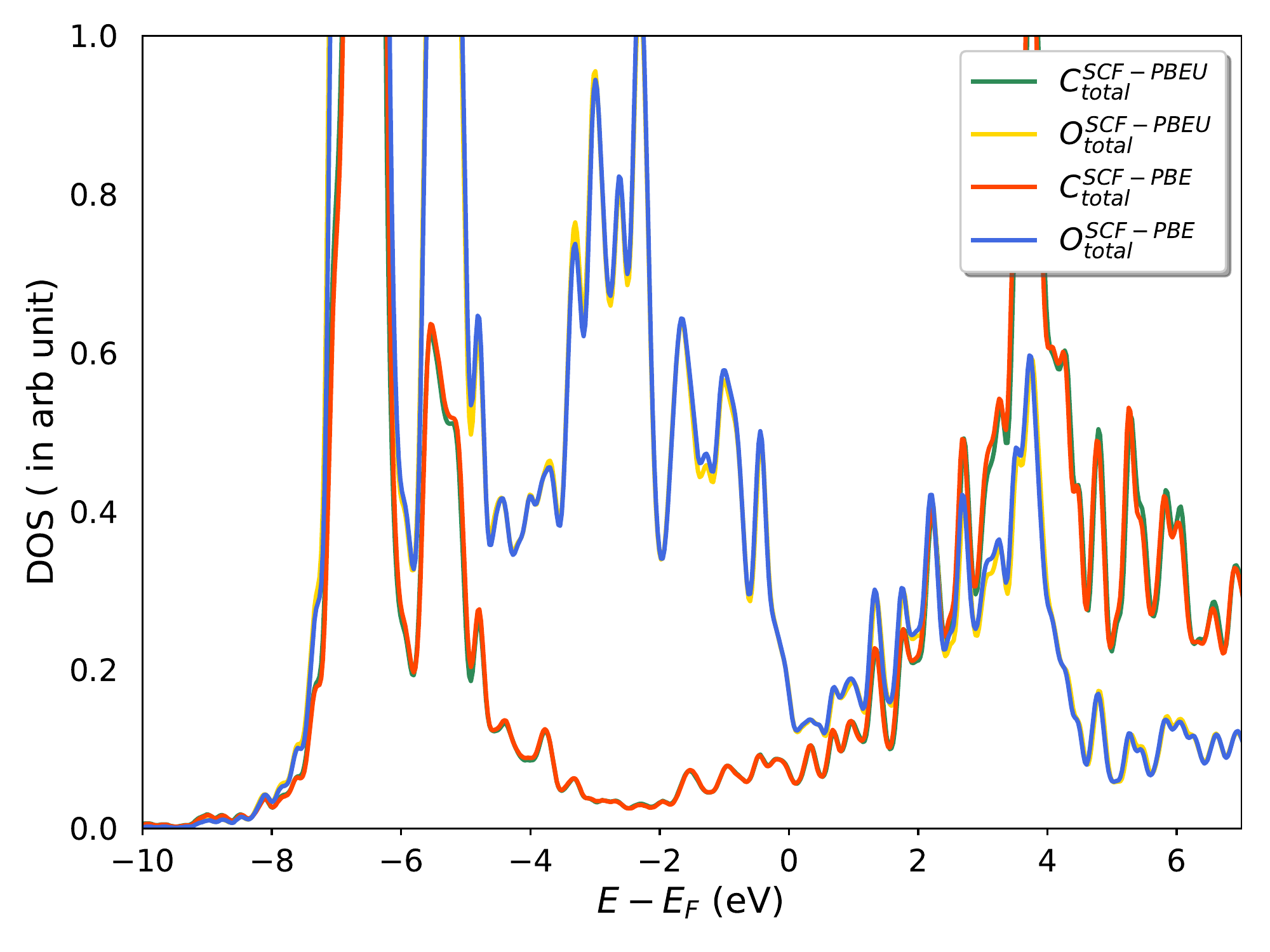}~
	\includegraphics[width=0.35\textwidth]{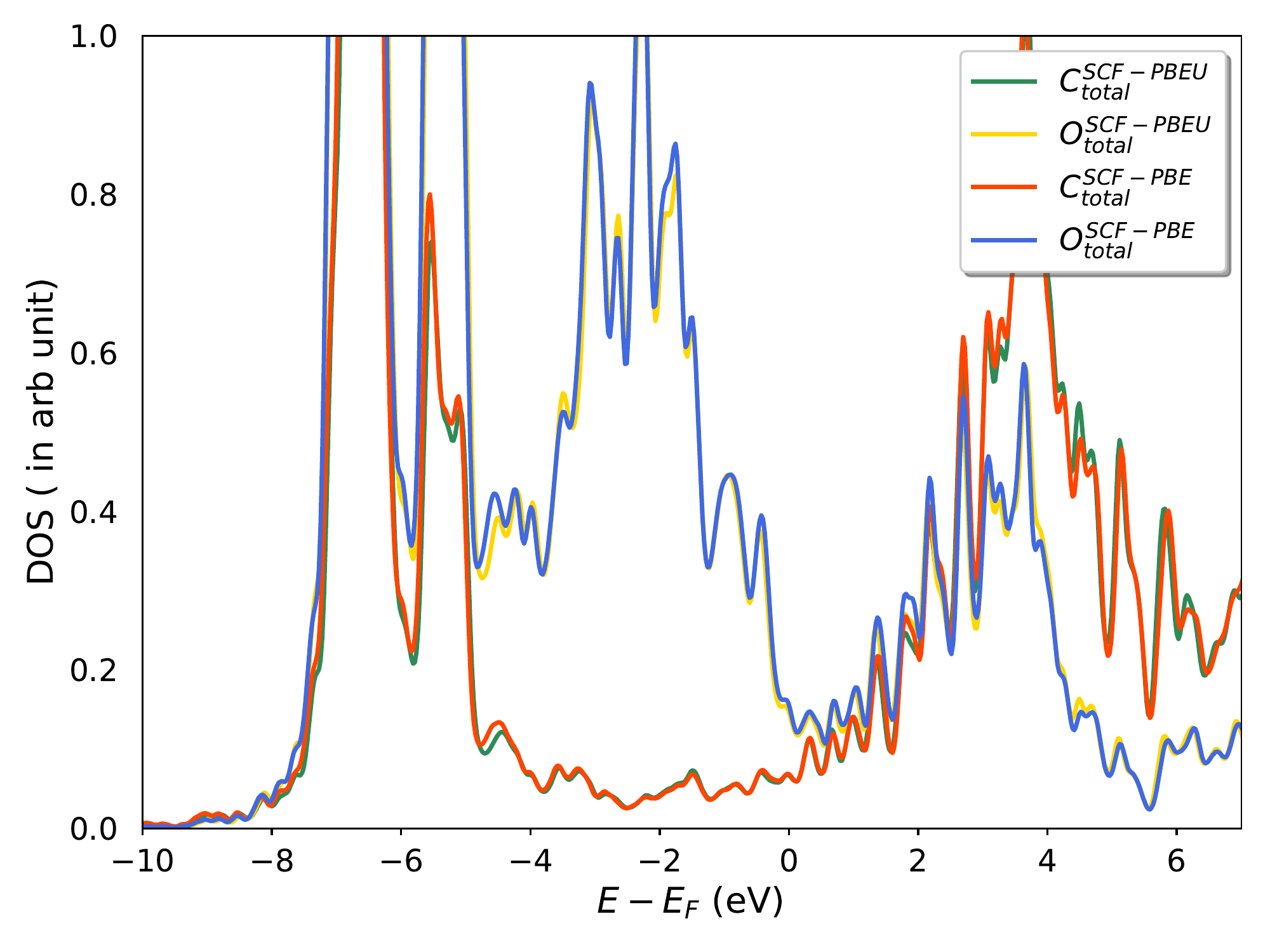}

	\caption{Atom-projected density of states (pDOS) of the adsorbed CO molecule in the `TOP', `FCC', and `HCP' configurations, computed using the selfconsistent PBE and PBE+U (with U = 0.75 eV) densities.}
%	\caption{pDOS of the adsorbed CO molecule for the ‘FCC’ configuration, computed using the selfconsistent PBE and PBE+U (with U = 0.75 eV) densities.}
	\label{fig:SI-dos-dcomposed-scfpbeu}
	
\end{figure*}

\vspace*{-7cm}

\FloatBarrier

%\bibliography{co-vdw-2017}{}
%\bibliographystyle{unsrtnat}